\documentclass[reprint, prl]{revtex4-2}
	\usepackage{graphicx}
	\usepackage{amsmath}
	\usepackage{amsfonts}
	\usepackage{amssymb}
	\usepackage{color}
	\usepackage{siunitx}
	\usepackage[colorlinks=true, citecolor=blue, urlcolor=blue, linkcolor=black]{hyperref}
	\usepackage{ragged2e}   %new code
\usepackage{dcolumn}% Align table columns on decimal point
\usepackage{bm}% bold math

\begin{document}

\preprint{APS/123-QED}
\title{The Effect of Parameter Variations on the Performance of the Josephson Travelling Wave Parametric Amplifiers}
	
\author{S.\,\'O Peat\'ain$^{1,2}$}\email{s.patton@lancaster.ac.uk}
\author{T.\,Dixon$^{2,3}$}
\author{P.\,J. Meeson$^3$}
\author{J.\,M. Williams$^2$}
\author{S. Kafanov$^1$}
\author{Yu.\,A. Pashkin$^1$}\email{y.pashkin@lancaster.ac.uk}
\affiliation{$^1$ Department of Physics, Lancaster University, Lancaster, LA1 4YB, United Kingdom}
\affiliation{$^2$ National Physical Laboratory, Teddington, TW11 0LW, United Kingdom}
\affiliation{$^3$ Department of Physics, Royal Holloway University of London, Egham, Surrey TW20 0EX, United Kingdom}	
\begin{abstract}
We have simulated the performance of the Josephson Travelling Wave Parametric Amplifier (JTWPA) based on the one-dimensional array of RF SQUIDs. Unlike the ideal model in which all SQUIDs are assumed to be identical, we allowed variation of the device parameters such as the geometric inductance of the SQUID loop, capacitance to ground, Josephson junction capacitance and critical current. Our simulations confirm the negative effects of variation of the device parameters leading to microwave reflections between individual cells and the shift of the flux bias from the optimal point. The strongest effect is caused by the variation of the geometric inductance as it varies both the wave impedance and the flux bias. The most detrimental, however, are point defects, such as shorts to ground making the circuit opaque to microwaves. This imposes stringent requirements on the fabrication process making it extremely challenging. We highlight the strict limitations on parameter spread in these devices while also discussing the robustness of the scheme to variation.

\end{abstract}
		
\maketitle
		
\section{Introduction}
Josephson weak links are ideal components for building parametric amplifiers due to their non-linear inductance and zero dissipation \cite{Barone}. Since the first realization of the Josephson parametric amplifier (JPA) \cite{Zimmer_1967} in the 1970s, enormous progress has been made in the development of JPAs \cite{Pedersen_1980}. These amplifier schemes operated in the four-wave \cite{Zimmer_1967, Parrish_1974, Feldman_1975, Wahlsten_1977} and three-wave \cite{Kanter_1971, Mygind_1978} mixing regimes, commonly referred to at that time as the four-photon and three-photon regimes (also doubly and singly degenerate modes) where the angular frequency of the signal, $\omega_{\mathrm s}$, the idler, $\omega_{\mathrm i}$, and the pump, $\omega_{\mathrm p}$, tones obey the relations $\omega_{\mathrm s} + \omega_{\mathrm i} = 2\omega_{\mathrm p}$ and $\omega_{\mathrm s} + \omega_{\mathrm i} = \omega_{\mathrm p}$, respectively. Internal pumping by the alternating Josephson current was used in some experiments \cite{Kanter_1971, Russer_1971, Kanter_1973, Kanter_1975, Vystavkin_1976, Thompson_1973}, however, external pumping using a microwave source \cite{Parrish_1974, Taur_1977} allows for full control of the pump frequency and power, hence better optimized device performance.

To increase interaction of the Josephson element with microwaves, a typical configuration of the JPA includes a resonant circuit coupled to the Josephson element, this imposes a limit on the amplifier's operating frequency and bandwidth set by the properties of the resonator (see, e.g., \cite{Zimmer_1967, Hatridge_2011}). To overcome this drawback, a travelling wave version of the JPA (JTWPA) was proposed \cite{Sweeny_1985} based on the analogy with the schemes used in quantum optics where a weak signal propagates through a non-linear optical medium driven by a strong pump tone. The JTWPA consists of a large number of series-connected junctions embedded in a transmission line, which can be regarded as a one-dimensional medium composed of artificial atoms with non-linearities arising from the current-phase relation of the Josephson junction.
%Despite being built of discrete building blocks, the medium can be treated as a continuum because the wavelength exceeds by far the atomic size. The JTWPA consists of a superconducting thin-film transmission line with a large number of embedded series-connected junctions (N$\sim\mathrm{10^3}$  with junction area A$_{JJ}$$\sim10^{-6}$). The long array of junctions embedded in the transmission line can be regarded as a one-dimensional  array of artificial atoms with non-linearities arising from the current-phase relation of the Josephson junction.  
This two-port, travelling-wave structure allows for broadband transmission up to its cut-off frequency with wave impedances close to 50\,\si{\ohm} throughout the range. The first successful realization of the JTWPA 
%consisted of a series array of 1000 Nb-based Josephson junctions embedded in a coplanar waveguide and 
demonstrated the gain of 16\,\si{\decibel} and the noise temperature of 0.5\,\si{\kelvin} at 1.7\,\si{\kelvin} \cite{Yurke_1996}, well below the noise temperature of even the most advanced HEMT amplifiers \cite{LNF_2018}.

\begin{figure*}
	\includegraphics[width=1\linewidth]{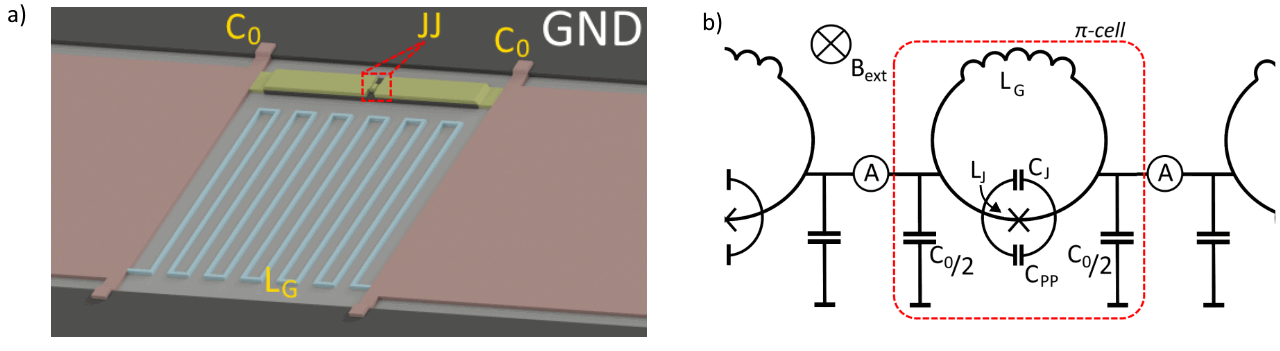}
	\caption{\label{Fig:CircuitDiagram}(a) An example JTWPA structure with electrical parameters defined by the denoted parts of the structure.  The loop with the geometric inductance, $L_G$ (blue), and the Josephson junction (yellow) and its associated inductance, $L_J$, the junction capacitance, $C_J$ and the  parallel plate shunt capacitance, $C_{PP}$ (grey) with the total junction capacitance denoted as $C_T$. The SQUID has the capacitance to the ground plane $C_0$ (pink). (b) The $\pi$-type cell model of the transmission line under study. The transmission line is approximated as a one-dimensional array of identical cells, each consisting of an rf-SQUID represented by lumped elements similar in principle to an LC ladder network. It is assumed in this model that there is no capacitive or inductive coupling between individual cells. A single unit cell is enclosed by the dashed line where A denotes the points in the array that data of propagating waves is taken. The whole structure is placed in a uniform magnetic field denoted as $B_{ext}$ to control the flux bias point. In our simulations we use the original parameter set suggested by Zorin for a 23\,\si{\ohm} matched line with $L_G=$57\,\si{\pico H}, $C_0 = $100\,\si{\femto F}, $C_T = C_J + C_{PP} = $ 60\,\si{\femto F} and $I_c=$5\,\si{\micro\ampere}.}
\end{figure*}

In recent years, superconducting parametric amplifier devices have seen renewed interest \cite{Aumentado:JTWPA}. The greatest motivation driving this development is the necessity for detecting extremely weak signals when reading out quantum circuits \cite{Bultink:Qubit_Readout, Macklin:RPM} or searching for dark matter \cite{Brubaker:Axions}. 
%The use of HEMT amplifiers alone for this purpose has proven unwieldy due to higher noise temperature \cite{LNF_2018} which has driven the development of the current JTWPA technologies. 
However, considerable problems exist within current JTWPA technologies including the cross and self-phase modulation effects, phase mismatch, pump depletion and higher order mixing tones, each of which hampers amplifier performance \cite{Zorin_2016}. Proposed solutions include resonant phase matching \cite{OBrien:RPM, Macklin:RPM, White:RPM}, quasi-phase matching \cite{Zorin:QPM}, and Floquet mode JTWPA's \cite{Peng_2021} although these techniques inherently limit the bandwidth of the device.
	
In this paper, we study a JTWPA scheme first proposed \cite{Zorin_2016} and further developed \cite{Zorin:QPM} by Zorin, which is sketched in Fig. \ref{Fig:CircuitDiagram}(a). Its lumped element model is shown in Fig. \ref{Fig:CircuitDiagram}(b). %This scheme has garnered much attention due to its robust and tunable operation. A realised scheme of this kind would be the ideal amplification device for many current experiments. 
The building blocks are unit cells enclosed by the dashed line consisting of rf-SQUIDs with parameters that set a wavelength of $\sim$40 nodes for a 5\,\si{\giga\hertz} tone in the medium, hence the lumped element model is suitable \cite{Walton_1987}. The application of the external magnetic field perpendicular to the SQUID loop plane allows the tuning of the non-linearity of the medium and operation of the amplifier in the much desired three-wave mixing regime. The proposed device was estimated to offer gain of $20\,\si{\decibel}$ in the bandwidth of $5.6\,\si{\giga\hertz}$ around a centre frequency of $6\,\si{\giga \hertz}$ \cite{Zorin_2016}. Further analysis of this scheme by Dixon \textit{et al.} \cite{Dixon_2020} showed that significant power is drawn to the higher harmonics causing a serious adverse affect on amplifier performance over any bandwidth. It was confirmed that the 3WM condition can be met, but that the tones considered should be extended in any analysis to include higher order mixing tones. 
%It is fair to say that this device has been well studied and is theoretically understood, however the lack of successful devices has tempered expectations and the goal of this paper is to investigate the viability of fabricating such a scheme at all.

% The key assumption of the design proposed in \cite{Zorin_2016} is that all the cells of the array are identical, ...	
The design proposed in \cite{Zorin_2016} assumes that all the cells of the array are identical. In reality, however, this is not possible even with the most advanced fabrication capabilities as variation in parameters is inherent regardless of the fabrication techniques. The main source of non-uniformity is in patterning where geometric variations produce varying inductances and capacitances. In addition to this, the oxidation process used to form the oxide barrier in the $\sim$10$^\mathrm{3}$ Josephson junctions is in the ballistic regime for each junction causing variation in the junction parameters of critical current (Josephson inductance) and junction capacitance, $I_\mathrm{c} \propto (1/L_\mathrm{J})$ and $C_\mathrm{J}$ respectively.  
%These variations will have a domino effect, most immediately on the impedance of the individual $\pi$-cells, then on the phase bias across the SQUIDs, then on the coupled mode equations that describe the amplification of the travelling wave and on many other things, all of which interact with each other leading to a complex system with potentially surprising characteristics.
The specific question we address in this paper is how variation of these device parameters affects the amplifier performance. It is expected that any departure from an array of perfectly similar rf-SQUIDs should lead to worsening performance, but the degree to which this is true, and variation of which parameter is most harmful in this regard, is unknown. We use Zorin's original set of parameters, as defined in Fig 1, as this device is already well understood.

%First, we start with the description of a analytical model of such an amplification array, to describe the properties of the device and the processes that occur within it \cite{Yaakobi:JPA}.
	
%\begin{figure}[bp]
%\includegraphics[width=1\linewidth]{circuit_diagram_new.png}
%\caption{\label{Fig:CircuitDiagram}The $\pi$-type cell model of the transmission line under study. The transmission line is approximated as a one-dimensional array of non-interacting identical cells, each consisting of an rf-SQUID represented by lumped elements: the loop with the geometric inductance, $L_G$, and the Josephson Junction and its associated inductance, $L_J$, the junction capacitance, $C_J$ and the  parallel plate shunt capacitance, $C_{PP}$. The SQUID has the capacitance to the ground plane $C_0$. One unit cell is enclosed by the dashed line. $n$ denotes the cell position along the array. The whole structure is placed in a uniform magnetic field denoted as $B_{ext}$ to control the flux bias point.}
%\end{figure}

	\section{Mathematical Model} \label{Math}
%		\begin{figure*}
%			\includegraphics[width=1\linewidth]{circuit_diagram.png}
%			\caption{\label{Fig:CircuitDiagram} A diagram of the rf-SQUID transmission line of the same kind described by Zorin \cite{Zorin_2016}. An extra capacitance has been added as many fabricated samples use a shunt capacitance across the junction in order to decrease the plasma frequency. Dimensions of an individual cell would be about 10\,$\mu m$ and so much smaller than the millimeter waves to be amplified allowing for a lumped element description of the device. An ladder type impedance model of a filter or transmission line. Such a 1D line acts as a low-pass filter with a cut-off frequency and in the special case where $Z_2$ is purely capacitive and $Z_1$ purely inductive all frequencies in the pass band will experience the same phase change over an individual cell.}
%		\end{figure*}
For clarity, we assume that the signal and pump tones propagate from left to right in the scheme shown in Fig.~\ref{Fig:CircuitDiagram}. Accordingly, the cells are also numbered from left to right such that $n = 0,1,2 ..., 1200$. The lumped element representation allows for the embedded rf-SQUID transmission line to be described by the $\pi$-cell model of a cascading filter or transmission line \cite{Walton_1987} shown in Fig. \ref{Fig:CircuitDiagram}(b). This scheme can be described to a large extent by the same relations used in the non-linear optics schemes to which it is related \cite{Boyd:Optics}.

In this model the array is described as a chain of two impedances, one impedance along the direction of propagation, $Z_1$ and one impedance to the return line, $Z_2$. In this case these are the impedance of the rf-SQUID and the impedance of the capacitance of the transmission line to the ground plane, respectively, this can be written in complex notation as:
		\begin{subequations}
			\begin{gather}
				Z_1 = \frac{1}{1/(i\omega L_G) + 1/(i\omega L_J) + i\omega C_T}\,\,, \\
				Z_2 = \frac{1}{i\omega C_0}\,\,,
			\end{gather}
			\label{Eq:Zs}
		\end{subequations}
\noindent where $\omega = 2\pi f$ and the other terms are as described in Fig. \ref{Fig:CircuitDiagram}.
In this way the impedance of an individual $\pi$-cell can be found to be:
		\begin{equation}
			Z_{\pi} = \frac{2Z_2\left(2Z_{\pi L}Z_2 + Z_1Z_{\pi L} + 2Z_1Z_2\right)}{4Z_{\pi L}Z_2 + 4Z_2^2 Z_1Z_{\pi L} + 2Z_1Z_2}\,\,,
			\label{Eq:pi_impedance}
		\end{equation}
\noindent where $Z_\pi$ depends on $Z_{\pi \textsc{L}}$, the (potentially different) impedance of the preceding cell.

For $Z_{\pi} \equiv Z_{\pi L}$, and approximating that $Z_1 = i\omega L_G$ as $i\omega L_G \ll i\omega L_J$ we can simplify Eq.\,(\ref{Eq:pi_impedance}) to:
		\begin{equation}
			Z_{\pi} = \sqrt{\dfrac{L_G}{C_0\left(1 - \dfrac{\omega^2}{\omega_c^2}\right)}}\,\,,
			\label{Eq:pi_impedance_simplified}
		\end{equation}

\noindent where the cut-off frequency of the array, $\omega_c$ is a function of the lumped elements and is given by:
		\begin{equation}
			\omega_c = \dfrac{2}{\sqrt{L_G \left(C_0 + 4C_T\right)}}\,\,.
			\label{Eq:cut-off-frequency}
		\end{equation}

\noindent From Eq\,(\ref{Eq:pi_impedance_simplified}) we see that for $\omega < \omega_c$, 
		\begin{equation}
			Z_{\pi} \approx \sqrt{\dfrac{L_G}{C_0}}\,\,.
			\label{Eq:characteristic_impedance}
		\end{equation}
It is evident from this that the parameters that will most greatly influence the impedance are $L_G$ and $C_0$. Junction capacitance/parallel plate shunt capacitance should have limited impact in this regard given the relative sizes of the impedances in this parameter set.

This approximation in Eq.\,(\ref{Eq:characteristic_impedance}) may not be valid with the implementation of dispersion engineering and a large junction capacitance. This would lead to a lower cut-off frequency and a greater phase-mismatch. Divergence from this and other assumptions may create significant barriers to correct operation hampering the device's usefulness. In this same regard the array must be terminated correctly to avoid any reflections from the end of the array and allow for proper power transmission.

Variation of the junction critical currents will have a limited affect on impedance as the Josephson inductance,

	\begin{equation}
		L_J = \frac{\Phi_0}{2\pi I_c}\frac{1}{\cos\varphi}\,\,,
		\label{Eq:Josephson-Inductance}
	\end{equation}
\noindent is theoretically infinite under our operating conditions. So the total cell inductance $L_{\mathrm{T}}^{-1} = L_J^{-1} + L_G^{-1}$ is dominated by $L_{\mathrm{G}}$ for the optimum three wave bias point $\varphi = \pi/2$. However, the optimal bias point itself,

	\begin{equation}
		\varphi_{dc} + \dfrac{2 \pi L_G I_c}{\Phi_0} \sin{\left(\frac{2\pi\phi_{dc}}{\Phi_0}\right)} = \varphi_{ext}\,\,,
		\label{Eq:transcendental}
	\end{equation}

\noindent may be perturbed by variations in $I_\mathrm{c}$. Which may impact the quadratic non-linearity coefficient,

	\begin{equation}
		\beta = \frac{2\pi L_G I_c}{\Phi_0}\sin\varphi_{DC}\,\,,
		\label{Eq:beta}
	\end{equation}
\noindent affecting the efficiency of the 3WM.

Apparently variation of critical current or geometric inductance will have an impact not only on the impedance of the rf-SQUID but also on the dominant wave mixing process and the maximum amplification that is possible in the device through the non-linearity, $\beta$.

Up to this point we have considered electrical parameters that can be controlled through device design. We should mention though that flux noise may have a significant effect on the device performance through the variation of the the phase across the Josephson junctions, $\varphi$, see Eqs.~(\ref{Eq:Josephson-Inductance}), (\ref{Eq:transcendental}) and (\ref{Eq:beta}). However, with proper magnetic shielding, flux noise can be suppressed to an insignificant level of about $1\,\si{\micro \Phi_0}/ \si{\hertz^{1/2}}$ \cite{Jabdaraghi_2017}.

Other things will also affect the phase, such as optical rectification \cite{Boyd:Optics} that leads to a drift in the phase bias over time. In addition, departure from the small pump approximation where $I_p \ll I_{DC}$ that has been proposed for optimal gain will cause a significant phase swing in the SQUIDs, potentially affecting the wave mixing dynamics in the device. In the following simulations, a pump amplitude of 0.985\,\si{\micro \ampere} at a frequency of 12\,\si{\giga \hertz} and a signal amplitude of 5\,\si{\nano \ampere} at a frequency of 7.2\,\si{\giga \hertz} have been chosen for continuity with \cite{Dixon_2020}.

\section{Simulation}
We complete simulations in the same manner as in Ref. \cite{Dixon_2020} beginning with an investigation of the impact of variation of electrical parameters of the SQUIDs and the transmission line on the performance of the device. Initially, we vary each parameter individually and then study the combined effect of variation of all electrical parameters simultaneously as would be the case in a fabricated scheme. Finally, we investigate the effects of different point defects that may occur in fabrication showing which are most devastating to device performance.
%We consider several possible standard deviations of device parameters variation, $\pm 1$\,\%, $\pm 5$\,\% and $\pm 10$\,\%. 

The ladder type model of a discrete element transmission line is simulated using the WRspice program, a platform proven adequate for use in Josephson circuits \cite{Dixon_2020, Shelly:SPICE, Naaman:JPA}. This program runs a nodal analysis of the circuit in the transient evolution, solving for the current and voltage at each node of the array using Kirchoff's laws. The Josephson junction is described using the resistively and capacitively shunted junction (RCSJ) model that can be fine tuned to match the characteristics of physical devices \cite{Whiteley_1991}. In this way we can simulate the propagation of waves through the array.

\begin{figure}[bp]
	\includegraphics[width=1\linewidth]{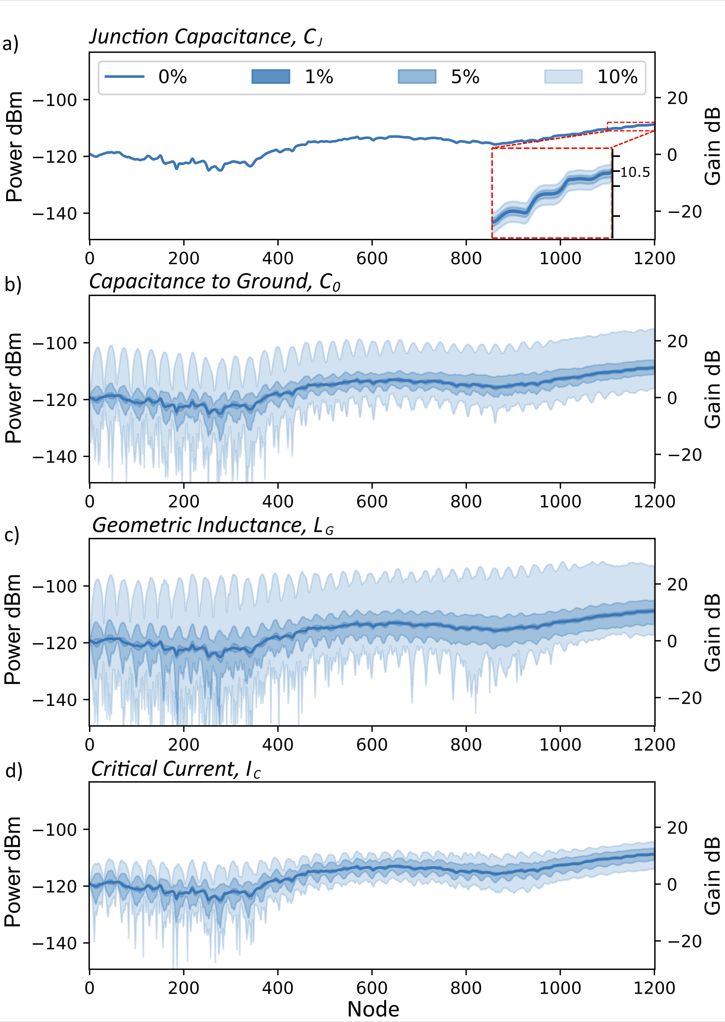}
	\caption{\label{Fig:Gaussian} Envelopes of signal wave gain caused by variation of various parameters. (a) Junction capacitance, $C_J$, has very little impact on the gain profile due to its limited effect on impedance or wave-mixing. (b) Capacitance to ground, $C_0$, creates a considerable envelope around the ideal signal profile mostly due to impedance mismatch. (c) Geometric inductance, $L_G$, has by far the greatest effect due to both the impedance mismatch, effect on flux bias and the wave mixing properties of the medium. (d) Critical current, $I_c \,\propto (1/L_J)$ has a considerable impact but less than either of $C_0$ (b) and $L_G$ (c). Its effect on impedance is minimal, but its large effect on the wave mixing properties compounds to create a sizeable envelope around the ideal case.}
	\end{figure}

An ideal scheme is first simulated with all the parameters equal throughout the SQUID array. The proper flux bias is calculated from Eq.\,(\ref{Eq:transcendental}) and inductively coupled to the geometric inductance of each rf-SQUID. The resulting signal amplification is shown as the solid lines in all panels of Fig.~\ref{Fig:Gaussian}.
% In these results the adverse effects of higher harmonics generation on the performance of the amplifier can be seen in the same way as \cite{Dixon_2020}. We build directly upon this work by introducing variation of parameters, likely to occur in fabrication of such a sample. 
Then we introduce a Gaussian distribution of the critical current of the junction, the shunt capacitance, the waveguide capacitance to ground, and the geometric inductance of the rf-SQUID in each $\pi$-cell of the transmission line. Each variation is done initially in isolation with standard deviations of $\pm1\%$, $\pm5\%$ and $\pm10\%$ to mimic the variation that may be expected from various fabrication techniques \cite{Naaman:JPA}.

In all systems where impedance mismatches occur, a ripple forms along the length of the wave due to complex interference between the forward and backward propagating waves having multiple reflections. This ripple has nodes along the length of the array for every $\lambda/2$ period of the signal wave, while the maximum of the signal gain between nodes depends on the particular wave mixing dynamics in the array \cite{Walton_1987}. In this device, a ripple is created for all occasions of variations of parameters which has implications on the gain and bandwidth when measuring the power output of such a device. The results shown in Fig.~\ref{Fig:Gaussian} display the possible final values for gain after a stochastic simulation method for 100 separate simulations.

	\begin{figure}[bp]
		\includegraphics[width=1\linewidth]{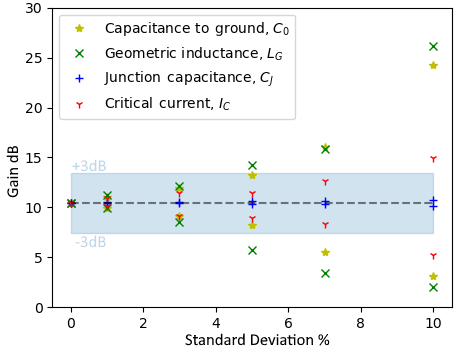}
		\caption{\label{Fig:Variation} The bounds of the envelopes from the stochastic simulations of variation in each parameter plotted against the respective standard deviation. Standard deviations that cause variation outside of the $\pm3\,\si{\decibel}$ band can be considered unacceptable as it would results in a narrow or inconsistent bandwidth.}
	\end{figure}

%It is assumed that each parameter is free to vary independently in this lumped element model and so the effects of variation in each can be studied in isolation.

	\begin{figure*}
		\includegraphics[width=1\linewidth]{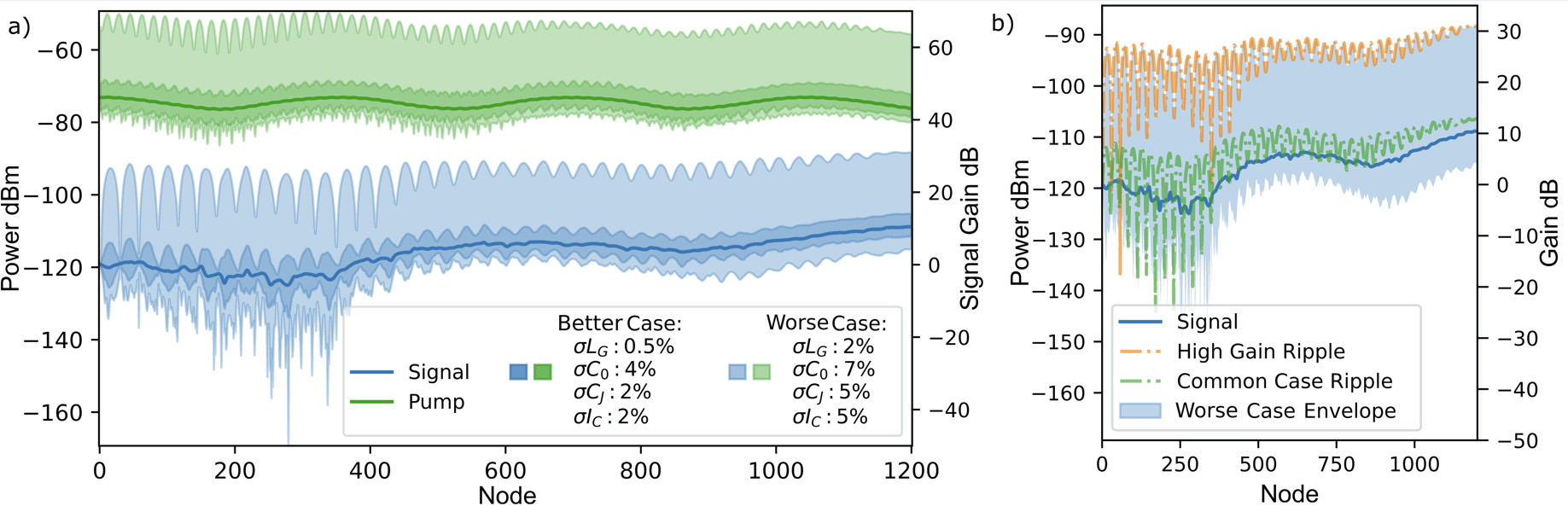}
		\caption{\label{Fig:Realistic_Variation}(a) Combined effect of variation of geometric inductance, capacitance to ground, junction capacitance and critical current, on the performance of the device with envelopes forming around both the signal and pump tones. Clearly, variations in impedance or proper biasing can seriously affect performance, even in the best possible case. (b) Two singular examples of the simulation results taken from the worse case simulations shown in their respective places in the envelope. The larger ripple (orange) and smaller ripple (green) are examples of the unpredictable effects that variation may have, with significantly higher gain possible under some conditions.}
	\end{figure*}

As seen in Eq.\,(\ref{Eq:pi_impedance_simplified}), junction capacitance has a minimal effect on impedance of the rf-SQUID or the flux biasing point. The small impact of variation in $C_J$ is clearly reflected in simulation results of Fig.~\ref{Fig:Gaussian}(a). We note, however, that should this parameter be changed significantly for the purpose of dispersion engineering or other endeavours to the point where $Z_{C_J} \sim Z_{L_G}$, then its impedance contribution is not negligible and we can assume variation in it would also incur a significant envelope.

The capacitance of the transmission line to the ground plane, $C_0$, enters the expression for impedance of the line, see Eq.\,(\ref{Eq:characteristic_impedance}), for our case where we consider $\omega < \omega_\mathrm{c}$, hence variation of this parameter causes significant variation of the wave impedance for our considered frequencies. The previously described ripple caused by this gives rise to considerable envelopes around the ideal case, shown in Fig.~\ref{Fig:Gaussian}(b).

As expected, the geometric inductance of the rf-SQUID, $L_G$ has the largest effect on gain than any other individual parameter, as shown in Fig. \ref{Fig:Gaussian}(c). As discussed in the previous section, this is due to the fact that geometric inductances determine the wave impedance and also impact the wave-mixing regime and strength of the 3WM non-linearity. One can see that variations in geometric inductance of $\pm 10$\% would produce devices with unpredictable properties, with the output gain falling in the range from almost zero up to about 25\,\si{\decibel}.

Variation in geometric inductance also directly affects the amount of flux penetrating the loop, therefore, in this way it can have a similar effect to what would be expected of flux noise. While the prominent ripple lends to the idea that impedance mismatches are the dominant factors impacting the device, it is however, difficult to tell how much of an effect the variation in flux has, or the effect of the strength of the non-linearity $\beta$ either.

The fact that this parameter is the most effectual is good in terms of device fabrication. The loop area can be made quite large to achieve the needed inductance, making the absolute variation that may occur a smaller percentage of the whole.

Variation of critical current can have a direct effect on the wave impedance of the line via the Josephson inductance, although its impact is likely to be small due to the large impedance of the Josephson inductance which is a direct consequence of the 3WM scheme proposed. Critical current also has a direct effect on the optimal flux biasing point and the strength of the 3WM non-linearity via Eqs.\,(\ref{Eq:transcendental}) and (\ref{Eq:beta}), respectively. It is not possible to simply separate the different contributions to the formation of the envelope in this case as all factors are inherently linked. However, it can be surmised from the approximations made in this model that the envelopes here are formed less from the impact of impedance mismatches and more greatly from the strength of the wave mixing processes themselves.

The difference in the impact of variation in the individual parameters can be studied through the bounds of the individual envelopes, this is plotted in Fig.\,\ref{Fig:Variation} which shows a  dependence of the envelope bounds on the standard deviation. We limit a useful device to output signal gain within $\pm 3$\,\si{\decibel} band of the ideal device as given by the blue shaded region.

The specific amount of variation that is tolerable to obtain a relatively flat bandwidth depends on the specific parameter set. However, this result shows that within some bounds ($I_c : \sigma=8\,\%,\,\, C_0: \sigma=4\,\%,\,\, L_G: \sigma= 3\,\%$) variation is tolerable for this parameter set at this frequency. This can further be tested through the simulation of a scheme with simultaneous variation in all parameters that is likely to occur in a realised sample.

%In most reasonable cases the final amplification is likely to lie within a particular uncertainty for any given frequency, to better understand this situation all parameters should be varied simultaneously so that their individual effects compound.

Using a realistic fabrication process we assume the percentage variation of each parameter for both a better and a worse case scenario, as shown in Fig.~\ref{Fig:Realistic_Variation}. The specific amount of random variation in parameters and any possible systematic variations across the chip greatly depend on the fabrication method used \cite{Naaman:JPA, Dolata_2005}. In this case we have again used the same Gaussian distribution of parameters with no spatial distribution.

From the envelopes it can be seen that the performance gets significantly worse with all added variations. However, the situation is reasonable with an optimal fabrication method and the gain of the amplifier at the final node can be expected to be 10 $\pm$ 2\,\si{\decibel}. Although, in the worse case studied the situation is considerably less promising with final node gain of 10 $^{+20}_{-10}$\,\si{\decibel}. This translates to an unpredictable gain in the worse case scenario.

	\begin{figure}[tp]
		\includegraphics[width=1\linewidth]{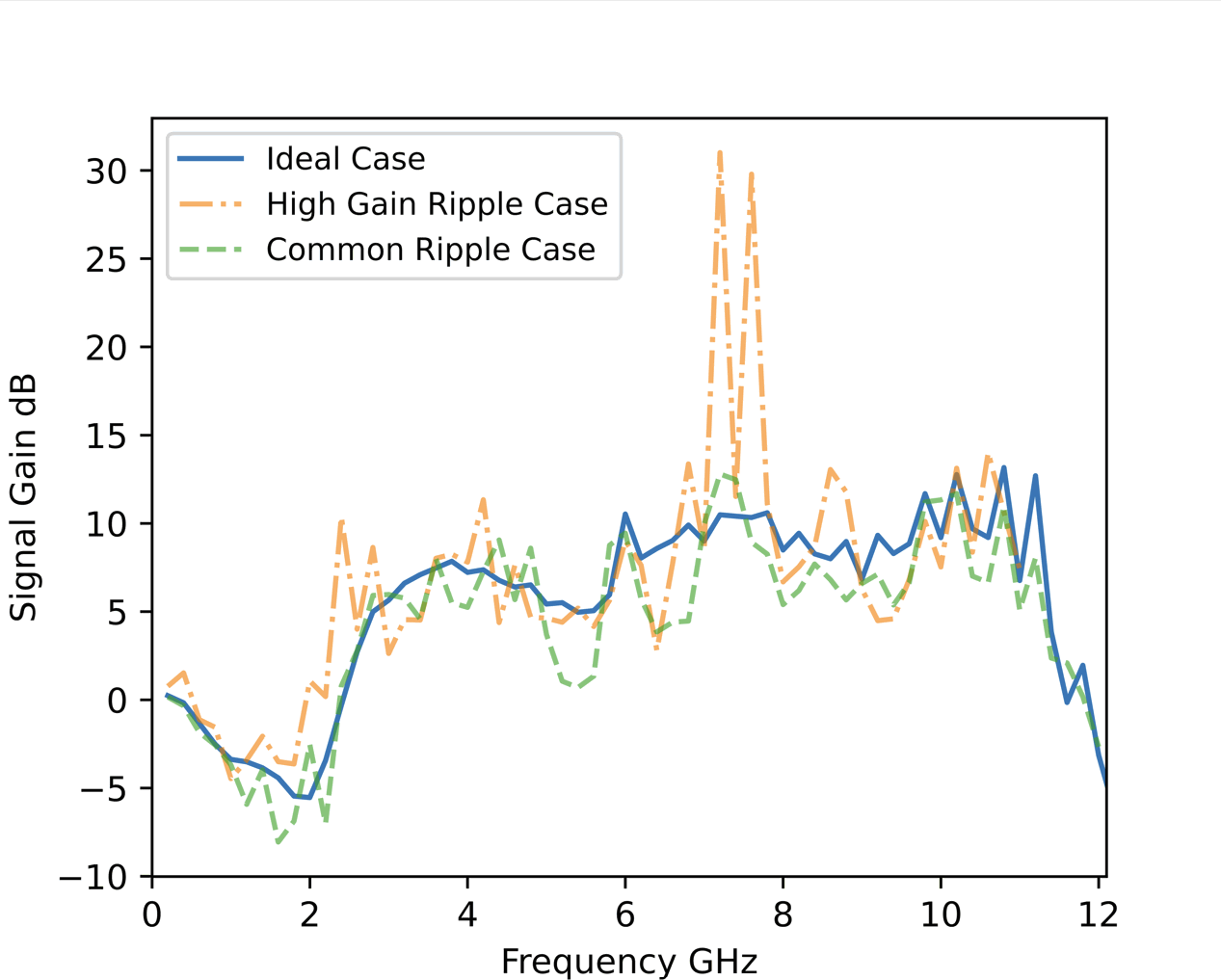}
		\caption{\label{Fig:Bandwidth} Plot of gain against signal tone frequency for a signal tone of 5\,\si{\nano \ampere} and a pump tone of 0.985\,\si{\micro \ampere} at 12\,\si{\giga \hertz}. The orange and green curves correspond to the parameter sets that create the similarly coloured curves in Fig \ref{Fig:Realistic_Variation}(b), while the solid line is the ideal case of no variations.}
	\end{figure}

It may be tempting to see possible silver-linings to these variations in that larger amplification can be seen in some conditions as shown in the case of two individual results taken from the make-up of the worse case envelope in Fig. \ref{Fig:Realistic_Variation}(b). This is shown in the bandwidth plot of these individual cases in Fig. \ref{Fig:Bandwidth}, where two spikes clearly emerge for the case of a ripple resulting in very high gain reaching 30\,\si{\decibel} at 7.2\,\si{\giga \hertz} (orange curves of Figs \ref{Fig:Realistic_Variation} and \ref{Fig:Bandwidth}). However, this effect is constrained to a specific signal frequency range likely due to some sort of resonance forming in the array. Such a device cannot be considered stable \cite{Macklin_2015, Planat_2020} and so generating high gain through these instabilities is impractical for a robust and reliable broadband amplifier. 

It should be noted though, that the device performance over the full frequency range for either of the cases with ripple does not differ too greatly from the ideal case with no variation, excepting the two obvious peaks. Apparently, a smaller ripple in the gain across the frequency response for each of the cases with variation may be hidden by the frequency step of $\sim$260\,\si{\mega \hertz} but we don't expect these ripples to be substantial. This is, in part, due to the fact that the size of the ripple is reduced towards the end of the array. The reason for the diminishing ripple is not currently properly understood but is thought to be a real physical effect due solely to the reflection characteristics of the device at the array ends \cite{Walton_1987}. This diminishing ripple shows how robust the device can be to variation of parameters and so, while a long array makes variations and defects more likely it can also correct for a certain amount of variation over its length.

	\begin{figure}[bp]
		\includegraphics[width=1\linewidth]{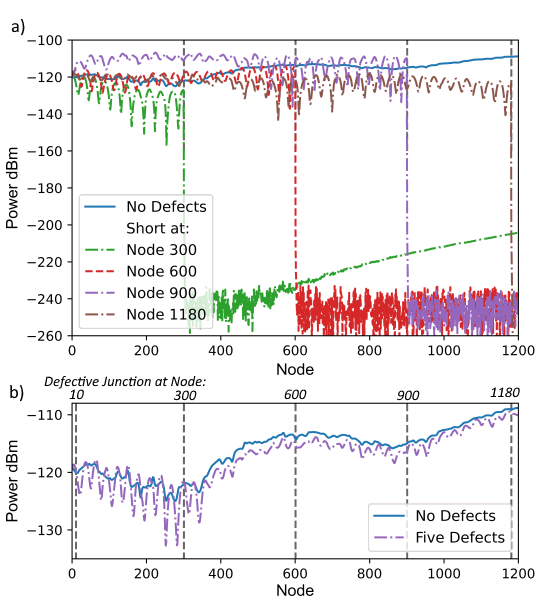}
		\caption{\label{Fig:Defects} Effect of point defects on the amplifier performance. (a) Shorts to ground due to broken $C_0$ at multiple places along the line causing  boundary reflections and the associated ripple prior to the short and attenuation of the signal to the simulation noise floor following the short. (b) A short across the Josephson junction is added at various points along the array. As shown the tones continue to propagate, but a significant impedance mismatch causes reflections from these points and a ripple forms along the gain curve.}
	\end{figure}

With modern techniques the variation that can be expected in the most consequential parameters like geometric inductance (loop area) can be kept low, allowing for functional devices to be fabricated. It is still possible that the variation that is natural to even the best techniques will result in aberrant behaviour at specific frequencies, however, the device may behave well aside from this.

Unfortunately, there is another common problem that may arise in fabrication, namely localised defects. Mistakes can occur in either fabrication or sample handling that lead to shorts along the line to the ground plane, broken inductive loops, or shorted junctions. These localised defects can have very different effects depending on the specific type of defect. This can include almost complete reflection of incident waves at the defect, or a small impedance change to the array depending on a shorted capacitor defect or a shorted junction defect respectively.

Perhaps, the most devastating defect to consider would be a short in the ground capacitor, the effects of which are shown in Fig. \ref{Fig:Defects}(a). Here we assumed that shorts occur at different places along the line, causing attenuation of the signal tone to the noise floor of the simulation from that point on in the line due to almost complete reflection at the impedance mismatch. Such reflections cause interference of the incoming and reflected waves resulting in ripples along the array before the short circuit.

Meanwhile a shorted junction is a defect commonly seen in long arrays of rf-SQUIDs such as this. This defect has limited effect on the propagation of waves through the array due to the geometric inductor and junction appearing in parallel in the rf-SQUID. The impact of the defects can compound so with several shorted junctions distributed along the array a ripple will begin to form that will affect the amplification as previously described. This situation is shown in Fig. \ref{Fig:Defects}(b) where the defects are marked by the vertical dashed lines. In this scheme the amplifier recovers over many nodes, thus the inclusion of the rf-SQUID provides a robustness to this type of typical defect allowing the device to still produce gain.

The situation is the same for a broken inductance loop if a flux bias is used. Although should a current bias be used instead, a broken geometric inductance may lead to $I>I_c$ in the junction, putting it in the dissipative regime.

Should the defective junction result in an open circuit as opposed to a short and there is also a defect in the geometric loop within a single cell then, no matter the biasing scheme, the step like change in propagation is again seen. It is, however, unlikely to see both these defects occur in a single cell.

\section{Conclusion}
In conclusion, we have completed a quantitative analysis of the effect of random parameter variations within the SQUID array proposed in reference \cite{Zorin_2016} on the JTWPA performance. The strongest effect is caused by the variation of geometric inductance as it affects both the impedance of the transmission line and flux bias for the 3WM regime in the rf-SQUID. However, the most devastating situation for the amplifier performance are the point defects such as shorts to ground whereas other defects within the rf-SQUID are not terminal but can compound if they occur in quantity diminishing the device performance. In such cases the impedance mismatch is considerable leading to strong reflections of microwaves and reduced gain.

We have also studied the maximum fabrication tolerances that can be present in such a device that will ensure the gain to be within $\pm 3$\,dB corridor. %While these tolerances will undoubtedly change for each parameter set, they are most likely to remain in a similar trend.
Finally, we have shown that even with a few point defects, shorted junctions, the device shows a robust performance, particularly for long arrays. Thus with state-of-the-art fabrication facilities the rf-SQUID embedded JTWPA is a strong candidate for broadband, high gain, high yield parametric amplifiers. Our approach to analysis of the parameter variation on the JTWPA performance is not limited to this scheme alone. Such analysis would give greater  confidence  in predicting  the behaviour of as yet unrealised samples.
%It is clear that these are important results not only for the development of three wave mixing JTWPA's but also those in the four wave mixing regime as it shows the robustness of this scheme to fabrication tolerances and will allow for greater confidence in predicting the behaviour of as yet unrealised samples.}

\section{Acknowledgements}
We thank C. Kissling, R. Dolata and A.B. Zorin for fruitful discussions and useful comments. This work was partially funded by the Joint Research Project 17FUN10 ParaWave of the EMPIR Programme co-financed by the Participating States and from the European Union’s Horizon 2020 research and innovation programme. We acknowledge support from the QSHS project ST/T006102/1 funded by STFC.
	\bibliographystyle{apsrev4-2}
	\bibliography{SGOP_bib}

\end{document}